\documentclass[preprint]{elsarticle}
\usepackage{ifpdf}
\usepackage[usenames]{color}
\usepackage{amssymb}
\usepackage{amsmath}
\usepackage{subfigure}
\usepackage{graphicx}
\usepackage{epstopdf}
\usepackage{subfigure}
\usepackage{color}
\begin{document}

%\runauthor{Fontanari and Serva}

\begin{frontmatter}

\title{Nonlinear group survival in Kimura's model for the evolution of
altruism}

\author[USP]{Jos\'e F. Fontanari\corref{cor1}}
\ead{fontanari@ifsc.usp.br}

\author[UFRN]{Maurizio Serva\fnref{fn2}}

\cortext[cor1]{Corresponding author. Tel.: +55 16 33739849; fax: +55 16 33739877}

\fntext[fn2]{On leave of absence from
Dipartimento di Ingegneria e Scienze 
dell'Informazione e Matematica, Universit\`a 
dell'Aquila,  Italy} 

\address[USP]{Instituto de F\'{\i}sica de S{\~a}o Carlos,
            Universidade de S{\~a}o Paulo,
            Caixa Postal 369, 13560-970 S\~ao Carlos SP, Brazil}

\address[UFRN]{Departamento de Biof\'{\i}sica e Farmacologia,
Universidade Federal do Rio Grande do Norte,
59072-970 Natal, RN, Brazil}

\begin{abstract}
Establishing the conditions  that guarantee the spreading or the sustenance of altruistic
traits in a population is  the  main goal of intergroup selection models. Of particular interest is the
balance of  the parameters associated to
group size,   migration and group survival   against the selective advantage of the non-altruistic individuals.
Here we  use Kimura's diffusion model of intergroup selection to  determine those conditions  in the case the
group survival rate is a nonlinear non-decreasing  function of the proportion of altruists in a group.
In the case this function is linear, there are two possible steady states which  correspond to  the non-altruistic
and the altruistic phases. At the discontinuous transition line separating these phases there is a non-ergodic 
coexistence phase.
For a  continuous concave survival function, we find an ergodic coexistence phase that occupies a finite region of
the parameter space in between the altruistic and the non-altruistic phases, and   is separated from  these phases by 
continuous transition lines. For a convex survival function, the coexistence phase disappears altogether but
a  bistable phase appears for which the choice of the initial condition determines whether the evolutionary dynamics leads
 to the altruistic or the non-altruistic steady state.
\end{abstract}

\end{frontmatter}

%
%----------------------------------------------------------------------------
\section{Introduction} \label{sec:Intro}
%----------------------------------------------------------------------------
%
The question of the evolution and maintenance of altruism or, as Wilson \cite{Wilson_75} put it  bluntly ``the surrender of personal genetic fitness for the enhancement of personal genetic fitness in others'', has been subject  of  stern dispute  since  the 1960s (see, e.g., \cite{Nowak_10,Simon_12a}).
The central point  of this debate is the potential of  intergroup selection, whose underlying  mechanism is the differential population (group) extinction,
to  counteract individual selection. Of particular historical relevance to this matter  was  Wynne-Edwards' suggestion that in order to control  population growth animals would limit their own fertility
for the sake of group survival \cite{Vero_62,Williams_66}.   In fact, if the extinction of groups occurs at a rate depending on
their composition, then such extinctions will, in principle, favor the existence of individuals that increase the probability of survival of the group they belong to.  The difficulty is
that the group extinction rates should have a magnitude comparable to that of individual  selection, a condition that, 
seemingly, lacks empirical support \cite{Wilson_75}.

The issue boils down  then to the identification of  the range of the parameters associated to the relevant evolutionary processes  
-- differential reproduction rate of individuals, differential extinction rate of groups,  migration and group size  (genetic drift) -- 
necessary to maintain an altruistic trait in the population.
Such a trait is defined as one that is detrimental to the fitness of the individual who expresses it, but that confers an advantage to the group of which that individual is a member. Hence the mathematical analyses  of the  large variety of group selection models for the evolution of altruism 
presented in the literature have provided the basic information  one needs to access the relevance of intergroup selection as an evolutionary
force in nature \cite{Boorman_80,Wilson_80}.
Moreover, the challenging mathematical models used to describe the resulting  two-level selection problem are viewed as an attraction on their own 
and have kept a recurrent theoretical interest on this controversial theory  
\cite{Eshel_72,Levin_74, Aoki_82,Kimura_83,Donato_97,Silva_99,Simon_10}.

In this paper we  offer exact  numerical and  approximate analytical solutions to perhaps the most elegant mathematical 
formulation of the intergroup selection problem proposed yet, namely,  Kimura's  diffusion model of intergroup selection \cite{Kimura_83}.
 The key quantity at the group selection level is the  group survival function $c \left ( x \right ) $ which essentially determines the
rate at which a group containing a fraction $x$ of altruists survives extinction.
Whereas Kimura  has considered the case that $c \left ( x \right ) $ increases
linearly with the frequency $x$  of altruists  within the group, i.e., $c \left ( x \right ) = c x  $, where $c$ is a positive
constant  (see also \cite{Ogura_87}), here we explore the  effects of an additional quadratic term, i.e.,
$c \left ( x \right ) = c x  + d x \left ( 1 - x \right) $, where $\mid d \mid \leq c$, whose effect  is to engender
convexity to the survival function.

The biological interpretation of the convexity of the survival function is similar to that of epistatic interactions between mutations or genes in genetics
\cite{Ridley_04}. In particular, in the linear case ($d=0$),  the beneficial effects that the altruists accrue to the group  are purely additive, i.e., they
 do not interact with each other. In the case of concave functions ($d>0$),  we have a situation akin to positive or synergistic epistasis in which
the group  benefit is greater  than the additive effects of the single altruists, whereas in the case of convex functions ($d<0$)
the  presence of the altruists together have a smaller effect than expected from their effects alone, a situation known in genetics as  negative or antagonistic epistasis. Hence the linear case studied by Kimura occurs when the individuals do not interact with each other in the group, which is
a somewhat unrealistic assumption since individuals within the same group should exhibit some sort of interaction.

In the case the group survival
 function is  concave ($d > 0$)  we find an ergodic coexistence phase in addition to the altruistic and non-altruistic  phases, whereas in the
case of a convex survival function  ($d < 0$) the coexistence phase is eliminated altogether but a bistable regime sets in for
sufficiently large  values of  $\mid d \mid $. Overall we conclude that a non-decreasing concave group survival function of the
frequency of altruist can increase significantly  the region in the parameter space where  the altruistic trait  can be maintained, albeit in combination
with the non-altruistic one.

The rest of the paper is organized as follows. In Section \ref{sec:model} we describe the evolutionary events that comprise the life cycles of the individuals
and groups, and derive the partial differential equation that governs the time evolution of
the proportion of  groups $\phi \left ( x , t \right )$ carrying a fraction $x$ of altruists at time $t$. In Section \ref{sec:linear} we re-examine 
the problem studied by Kimura, $c \left ( x \right ) = c x  $, and offer rigorous arguments to locate the transition line between the altruistic and non-altruistic phases as
well as to characterize the non-ergodic coexistence regime at the transition line. In Section \ref{sec:quad} we consider the
effect of adding the  quadratic term  $d x \left ( 1 - x \right) $ to  the linear term considered by Kimura. The resulting  phase-diagram is
studied in detail for the concave case ($d > 0$) and the continuous transition lines separating the ergodic coexistence phase from the
altruistic and non-altruistic phases are determined numerically and analytically.
 Our arguments to prove the existence of a bistable  regime in the case $d<0$ are presented in the
Appendix. In  Section \ref{sec:Heav} we describe succinctly the results for
a Heaviside survival function and show that the continuous transition between the non-altruistic and coexistence phases 
observed in the case of a concave survival function becomes discontinuous for the Heaviside function.
Finally, in Section \ref{sec:conc} we summarize our main results and present some concluding remarks.

%
%----------------------------------------------------------------------------
\section{The model}\label{sec:model}
%----------------------------------------------------------------------------
%

We consider a meta-population composed of an infinite number of  competing groups. Each group encompasses
exactly $N$ haploid, asexually reproducing individuals. There are two  alleles at
a single locus that determine whether an individual is altruist (allele $A$)  or not (allele $B$). The
fitness of the individuals are fixed solely by this trait  - altruists are assigned fitness 1 and   non-altruists
fitness $1+s$, where $s \geq 0$ is a parameter on the order of $1/N$. We assume that 
$N$ is sufficiently large so that the frequency of  altruists  within a group, denoted by  $x$,
can be viewed as a continuous variable in the closed interval $ \left [ 0,1 \right ]$. The meta-population   is
described by the proportion of groups $\phi \left ( x, t \right ) \Delta x$  whose frequency of  altruists lies in the range $\left ( x , x + \Delta x \right )$  at time $t$.
Our goal is to determine  how the probability density $\phi$ is affected
by the three  evolutionary processes: the individual  competition within a group, the  migration of individuals between groups
and the competition between groups.  In the following we discuss  these processes  in detail. 

%----------------------------------------------------------------------------
\subsection{Individual selection}
%----------------------------------------------------------------------------

If we assume that a group contains $j$ altruists (hence $N-j$ non-altruists), then the probability that  there will be exactly $i$ altruists after
reproduction is given by the  Wright-Fisher process
\cite{Crow_70}
\begin{equation}\label{rij}
r_{ij} = \left( {\begin{array}{*{20}c} N \\ i \\ \end{array}} \right) w_j^i \left ( 1 - w_j \right )^{N-i}
\end{equation}
where $w_j = j/\left [ N + s \left ( N - j \right ) \right ]$ is the relative fitness of the subpopulation of
altruists in the group. 
The way this process affects the probability density $\phi \left ( x, t \right )$ is  derived  using the  diffusion 
approximation of population genetics, which consists essentially on the calculation of
the jump moments $\left \langle \left ( x' - x \right ) \right \rangle_r$ and $\left \langle \left ( x' - x \right )^2 \right \rangle_r$
where $x = j/N$ and $x'= i/N$  are the frequencies of altruists before and after  reproduction, respectively. Here
$\langle \ldots \rangle_r$ stands for an average using the transition probability $r_{ij}$. These moments
contribute to the  drift  and the diffusion  terms of a Fokker-Planck-like equation  for $\phi$ (see eq.\ (\ref{eqphi})). We refer the reader to
\cite{Baxter_07,Blythe_07}  for a detailed discussion of the diffusion approximation.
More pointedly,
direct evaluation of the jump moments using the transition probability (\ref{rij}) yields 
\begin{equation}\label{d1r}
\left \langle \left ( x' - x \right ) \right \rangle_r = w_j - x \approx - s x \left ( 1 - x \right )  
\end{equation}
and 
\begin{equation}\label{d2r}
\left \langle \left ( x' - x \right )^2 \right \rangle_r =  \frac{1}{N} w_j \left ( 1 - w_j \right ) + \left ( w_j - x \right )^2  \approx \frac{1}{N}  x \left ( 1 - x \right )  
\end{equation}
where  we have kept only terms of the first order in  $1/N$ 
(recall that the fitness advantage $s$ of the non-altruists 
is on the order of $1/N$).

%----------------------------------------------------------------------------
\subsection{Migration}
%----------------------------------------------------------------------------

Following  Wright's island  model we  assume  that  $J$ individuals of each group are replaced by migrants
in the time interval $\Delta t$  and that the frequency of altruists among the migrants is  equal to the average frequency
of altruists in the entire meta-population, i.e., $\bar{x} = \int_0^1 x \phi \left ( x, t \right ) dx$ \cite{Wright_51}.  The probability
that a group with $j$ altruists ($x = j/N$) becomes a group with $i$ altruists ($x' = i/N$) due to the migration process is then
\cite{Aoki_82}
\begin{equation}
m_{ij} = \sum_{k=k_l}^{k_u} \frac{ \left( {\begin{array}{c} j \\ k \\ \end{array}} \right) \left( {\begin{array}{c} N -j\\ J-k \\ \end{array}} \right)}
{\left( {\begin{array}{c} N \\ J \\ \end{array}} \right)}  \left( {\begin{array}{c} J \\ i -j + k\\ \end{array}} \right) 
\bar{x}^{i-j+k} \left ( 1 - \bar{x} \right )^{J-i+j-k}
\end{equation}
where $k_l = \max \left (j-i,0,J-N+j\right)$ and $k_u = \min \left ( j,J-i+j,J \right )$. 
This somewhat formidable expression has a simple interpretation: the hyper-geometric component  yields the probability that exactly 
$k$ altruists and $J - k$ non-altruists are eliminated from the group to make room for the $J$ migrants, whereas the binomial part yields the probability that
there are exactly $i-j+k$ altruists among the $J$ migrants. Note that after migration the number of altruists in the group is given by the sum of 
the altruist originally in the group  $\left ( j-k \right )$  and the number of altruists among the migrants $\left ( i-j+k \right )$.
The first two jump moments are given by
\begin{equation}\label{d1m}
\left \langle \left ( x' - x \right ) \right \rangle_m = m \left ( \bar{x} - x \right ) 
\end{equation}
and 
\begin{equation}\label{d2m}
\left \langle \left ( x' - x \right )^2 \right \rangle_m=  \frac{m}{N} \bar{x} \left ( 1 - \bar{x} \right ) + m^2 \left ( \bar{x}- x \right )^2  
+ \frac{m \left ( 1 - m \right )}{N-1}  x \left (1 - x \right )
\end{equation}
where $\langle \ldots \rangle_m$ stands for an average using the transition probability $m_{ij}$ and 
 $m = J/N$ is the fraction of the local population that is replaced by migrants. Assuming that $m$ is on the  order of
$1/N$, i.e., that the number of migrants $J$  remains finite and limited when $N$ grows large,  we can
neglect the second jump moment which is $O \left ( 1/N^2 \right )$. In addition, the first jump moment  becomes of the same order of the 
drift contribution due to the selective advantage of the non-altruists, eq.\ (\ref{d1r}).

%----------------------------------------------------------------------------
\subsection{Intergroup selection}
%----------------------------------------------------------------------------

Since $x$ represents the fraction of altruists in a group,  we define the group survival   rate $c \left ( x \right ) $
as a monotone non-decreasing function of $x$. Assuming that a proportion
$1 - \left [ a -  c \left ( x \right ) \right ] \Delta t$ of the  groups 
carrying the fraction $x$ of altruists survives extinction during time interval
$\Delta t$ we can write
\begin{equation}
\phi \left ( x , t + \Delta t \right ) =  \left [ 1 -  \left [ a -  c \left ( x \right ) \right ] \Delta t \right ] \phi \left ( x , t \right )  \zeta
\end{equation}
where $a$ is some arbitrary rate. Here $\zeta$ is such that $\int_0^1 dx \phi \left ( x , t + \Delta t \right )  =1$, i.e, 
$\zeta = 1/\left [ 1 - \left ( a - \bar{c}\right ) \Delta t \right ]$ with
\begin{equation}\label{cbar}
\bar{c} = \int_0^1 c \left ( x \right ) \phi \left ( x , t  \right )  dx.
\end{equation}
The enforcement of  the normalization of $\phi$ after the extinction process is akin to assume that
the extinct groups are recolonized or replaced by the surviving ones in proportion to their frequencies. 
Finally, taking the limit $\Delta t \to 0$ we obtain the change in $\phi$  due to the extinction and
recolonization processes, 
\begin{equation}\label{IG}
\Delta \phi = \left [ c \left ( x \right ) - \bar{c} \right ]  \phi \left ( x , t  \right ) \Delta t ,
\end{equation}
from where we can see that  the arbitrary rate $a$ has no effect at all on the intergroup selection process.
Equation (\ref{IG}) implies that the process of extinction followed by recolonization amounts to an effective competitive interaction between
groups.

For the most part of the  paper, we will focus on the group  survival function 
\begin{equation}\label{cx}
c \left ( x \right ) = c x + d x \left ( 1 - x \right )
\end{equation}
with $ c > 0$.  
Clearly, $c \left ( x \right )$ is non-decreasing in the interval $\left [ 0,1 \right ]$
provided that  $c \geq \mid d \mid $, in which case the model can be said to 
 describe the  competition between 
individual selection favoring non-altruistic individuals ($s>0$) and intergroup selection 
favoring altruistic individuals. In addition,
$c \left (0 \right )=0$ and $c \left ( 1 \right )=c $ regardless of the value of $d$.  

We note that  the sole role
of $d$ in eq.\  (\ref{cx}) is to generate convexity. The case $d=0$ was studied by Kimura  in the
context of the evolution of an altruistic character \cite{Kimura_83}, whereas we have recently
considered the  case $c=0$ (and $m=0$ as well) in a prebiotic evolution 
scenario for  the coexistence of self-replicating molecules  \cite{Fontanari_13}. 
The motivation behind prescription (\ref{cx}) is to understand the effect of a weak nonlinearity,  modeled here by  the quadratic
term $-d x^2$,  on the linear case studied by Kimura. In addition, in Section \ref{sec:Heav} we will consider briefly  the effect of a strong nonlinearity  
where $c \left ( x \right )$ is given by  a Heaviside  function.

%----------------------------------------------------------------------------
\subsection{Evolution equation for $\phi \left ( x, t \right )$}
%----------------------------------------------------------------------------

Combining the changes in $\phi = \phi \left ( x, t \right )$  due to the three processes described above  we obtain
\cite{Kimura_83}
\begin{equation}\label{eqphi}
\frac{\partial }{\partial t} \phi = \frac{1}{2N} \frac{\partial^2}{\partial x^2} \left[ x \left (1-x \right )\phi  \right]
-\frac{\partial}{\partial x} \left[ a \left (x, t \right ) \, \phi \right]
+ \left  [ c \left (x \right )-\bar{c} \left (t \right ) \right ] \phi 
\end{equation}
where   
\begin{equation}\label{drift}
a \left (x,t \right ) = -s x \left (1-x \right ) - m \left [ x- \bar{x} \left ( t \right ) \right ]
\end{equation}
is the drift  term,
\begin{equation}\label{def_xt}
\bar{x} \left ( t \right )=\int_0^1 x \, \phi \left (x,t \right ) \, dx 
\end{equation}
is the mean frequency of altruists in the  meta-population,
and 
\begin{equation}\label{def_ct}
\bar{c} \left ( t \right ) =  \int_0^1 c \left ( x \right ) \, \phi \left (x,t \right )\, dx .
\end{equation}
is  the mean group survival  rate. In addition, the normalization condition 
$\int_0^1 \phi \left (x,t \right ) \, dx = 1$ holds for all times $t$.

We note that in the deterministic
limit $N \to \infty$ the diffusion term of eq.\ (\ref{eqphi}) can be neglected and Kimura's 
partial differential equation  reduces to a particularly simple realization of the deterministic
model of group selection studied in  \cite{Simon_10}.

%----------------------------------------------------------------------------
\subsection{Equation for the steady state}
%----------------------------------------------------------------------------

In the limit $t \to \infty$ the system reaches equilibrium  so that $\partial \phi/\partial t = 0$ 
and the steady-state equilibrium probability density $\phi \left (x,t \right ) \to \hat{\phi}\left (x \right )= \hat{\phi}$ satisfies
\begin{equation}\label{eqeq}
\frac{\partial^2}{\partial x^2} \left[ x \left ( 1-x \right ) \hat{\phi}  \right]
+\frac{\partial}{\partial x}
\left[ A \left (x \right  ) \hat{\phi} \right]
+\left  [ C \left (x \right )-\bar{C} \right ] \hat{\phi} = 0
\end{equation}
where  $A \left ( x \right ) =  S x \left (1-x \right ) + M \left ( x-\bar{x} \right ) $ and we have introduced
the rescaled parameters $S = 2 N s$ and $M = 2 N m$, as well as the  rescaled survival rate
$C \left ( x \right ) = 2N c \left ( x \right )$.  In addition,
\begin{equation}\label{barxc}
\bar{x}=\int_0^1 x \, \hat{\phi} \left ( x \right )  \, dx \; \; \; \; \; \; \; \; \;\;
\; \; \; \;  \bar{C}=\int_0^1 C \left ( x \right ) \, \hat{\phi} \left ( x \right )  \, dx .
\end{equation}

For $M > 0$, eqs.\ (\ref{eqeq})  and (\ref{barxc}) are satisfied both by $\hat{\phi} = \delta \left ( x \right )$ and    
$\hat{\phi} = \delta \left ( x-1 \right )$, and they may also  be
satisfied by a regular function $\hat{\phi} = \hat{\phi}_r \left ( x \right )$. 
However, the migration term prohibits solutions that are combinations of these three possibilities, since in this case
those equations  are violated in at least one of the two extremes, $x=0$ or $x=1$. 
In other words, 
\begin{equation}\label{com}
\hat{\phi} \left (x \right ) =A_0 \delta \left ( x \right ) + B \hat{\phi}_r \left ( x \right )   + A_1 \delta \left ( x - 1\right ) 
\end{equation}
can be a solution only if one of the three coefficients $A_0$, $B$ or $A_1$
equals one and the other two equal zero.  As a result, there are three potential phases at
the steady state: a non-altruistic phase $\hat \phi = \delta \left ( x \right )$, an altruistic phase
$\hat \phi = \delta \left ( 1-x \right )$, and a coexistence phase
$\hat{\phi} = \hat{\phi}_r \left ( x \right )$ where the two individual types cohabit  a same group.

However, the linear combination (\ref{com})  with two or three non-vanishing coefficients  is a solution 
of eq.\ (\ref{eqeq}) in
the case of isolated groups, $M=0$ \cite{Fontanari_13}. This situation is useful to elucidate the nature of the averages involved
in the derivation of eq.\ (\ref{eqeq}). In fact, because the number of groups is infinite, stochasticity occurs only in
the processes that take place inside the groups. For example, in the absence of group selection (i.e., $c\left ( x \right ) = 0$
for $x \in \left [0,1 \right ]$),
each group represents an independent realization of the Wright-Fisher process and  
since in this case the intragroup dynamics leads to the fixation of one of the  individuals types in the group
we can interpret $A_0$ in eq.\ (\ref{com})  either as the fraction of groups in which occurred fixation of the egoistic type 
or as the probability that the egoistic type  fixates in a given group. Hence the metapopulation, which is composed of $A_0$ purely egoistic
and $A_1$ purely altruistic groups, is the ensemble of the
realizations of the Wright-Fisher process. A similar interpretation holds in the presence of group selection $c\left ( x \right ) > 0$,
except that the groups are no longer  independent in this case which results in a biased ensemble of the intragroup stochastic
process.

It is instructive to mention that if a regular solution exists,  then integration of eq.\ (\ref{eqeq}) over the interval $\left [ 0, 1 \right ]$
yields
\begin{equation}
\left. \frac{d}{dx} \left ( x \hat{\phi}_r \right ) - M \bar{x} \hat{\phi}_r \right |_{x=0} = 0 
\end{equation}
and
\begin{equation}
\left. \frac{d}{dx} \left [ \left (1- x \right )\hat{\phi}_r \right ] + M \left ( 1 - \bar{x} \right ) \hat{\phi}_r \right |_{x=1} = 0
\end{equation}
which imply that for $x$ close to $0$ one has $\hat{\phi}_r \sim x^{M \bar{x} -1}$, whereas for $x$ close to $1$ one has
$\hat{\phi}_r \sim \left ( 1 - x \right )^{M \left ( 1-\bar{x}\right ) -1}$. 

%
%----------------------------------------------------------------------------
\section{Linear  group survival  rate}\label{sec:linear}
%----------------------------------------------------------------------------
%

This is the case considered in the seminal paper of Kimura  \cite{Kimura_83} and corresponds to the choice $d =0 $ in
eq.\ (\ref{cx}) so that $C(x)=C x$ with $C = 2N c $. A simplifying feature of the linear case is that the dynamical variable (\ref{def_ct})
becomes  $\bar{C} \left ( t \right ) = C\bar{x} \left ( t \right )$ and so eq.\  (\ref{eqphi})
exhibits  only one non-local dynamical variable, namely, $\bar{x} \left ( t \right )$.  Here we offer a much simpler approach than
that presented by Kimura, which does not involve the numerical solution of the steady-state equation. Kimura's approach 
 was based on the  presence of a small mutation rate between
the alleles $A$ and $B$ which guarantees the existence of a regular solution for all values of the model parameters.

We begin by multiplying  both sides of eq.\ (\ref{eqphi}) 
by $ e^{Cx/M} $ and integrating over the interval 
$\left [0,1 \right ]$, which yield
\begin{equation}\label{kim}
\frac{\partial}{\partial \tau} \int_0^1 e^{Cx/M}  \phi  \,dx    = 
 \frac{C R}{M^2}
\int_0^1 e^{Cx/M}  x \left ( 1 - x \right ) \phi   \,dx 
\end{equation}
where $\tau = t/2N$ and $R \equiv C - MS$. 

If we assume that  $R \neq 0$, then 
the right hand side of (\ref{kim}) must equal zero at the steady state $\phi \left (x,\tau \right ) \to \hat{\phi} \left (x\right ) $.
Since $\int_0^1 e^{Cx/M}  x \left ( 1 - x \right ) \hat{\phi} \, dx $ is strictly
positive unless $\hat{\phi}=\delta \left ( x \right )$ or $\hat{\phi}=\delta \left (1-x \right)$, we must conclude that only these
two singular steady-state solutions are allowed. Next, let us assume that eq.\ (\ref{kim}) holds with $R > 0$.
Then the right hand side is always positive (provided the initial distribution  $\phi \left (x,0 \right )$ is not a  Dirac delta 
centered at $0$ or $1$), and so 
 $ \int_0^1 e^{Cx/M} \phi  \,dx$ 
increases until it reaches,   for $\tau \to \infty$,  the maximum value 
$e^{C/M}$ which implies that $\hat{\phi} = \delta \left (x-1 \right )$.
Analogously, assuming that eq.\  (\ref{kim}) holds with $R < 0$
the same reasoning leads to the 
conclusion that $ \int_0^1 e^{Cx/M}  \phi \,dx$ decreases with increasing $\tau$ until it 
reaches the minimum value $1$, which entails that   $\hat{\phi} = \delta \left (x \right )$.
In sum,  eq.\ (\ref{kim})  implies that  
$\phi \left ( x,\tau \right ) \to \delta \left (x-1 \right )$ for $R > 0$ 
(provided the initial condition is not $\phi(x,0) = \delta(x)$)
and that
$\phi \left ( x,\tau \right ) \to \delta \left (x \right )$ for $R < 0$
(provided the initial condition is not $\phi(x,0) = \delta(x-1)$).

It  remains to analyze the model at the  transition surface $R = 0$. In this case 
we can easily  verify that 
the steady-state solution of eq.\ (\ref{eqeq}) is the Beta distribution
\begin{equation}\label{sol}
\hat{\phi}_{k} \left ( x \right )= k \, x^{M \bar{x}-1} \,  (1-x)^{M (1-\bar{x})-1} 
\end{equation}
where the normalization factor $k$  is the reciprocal of the standard Beta function, i.e.,
$k= 1/B \left [ M \bar{x},  M \left (1-\bar{x} \right )  \right ]$. Since  eqs.\  (\ref{barxc})  are satisfied for any choice of
$\bar{x}$, the value of this parameter must be determined by the initial distribution $\phi \left ( x, 0 \right)$. In fact, setting
 $S = C/M$  in eq.\ (\ref{kim}) we find that
$\int_0^1 e^{S  x}\phi (x,t)\,dx$ is constant in time  and so
\begin{equation}\label{cons}
\int_0^1 e^{S  x}\phi \left ( x,0 \right ) \,dx   =
\int_0^1 e^{S  x} \hat{\phi}_k \left ( x \right ) \,dx ,
\end{equation}
which provides the necessary condition to determine $\bar{x}$ univocally from
the knowledge of the initial distribution $\phi \left ( x, 0 \right)$ and parameters $C$ and $M$.
For example, in the limit $S \to 0$  eq.\ (\ref{cons}) yields $\bar{x} = \int_0^1 x \phi \left ( x,0 \right ) \,dx  $,
i.e., $\bar{x}$ is a constant of movement in this case.

In conclusion, in the case of the linear survival rate we have three steady-state phases: a non-altruistic  phase for $R = C-MS < 0$,
an altruistic phase  for $R = C -MS > 0$ and a non-ergodic coexistence phase
at the transition surface $R=0$.

%
%----------------------------------------------------------------------------
\section{Quadratic  group survival rate}\label{sec:quad}
%----------------------------------------------------------------------------
%

Here we consider the complete prescription (\ref{cx}) for group survival, which is written in terms of the rescaled parameters as
\begin{equation}\label{quad}
 C \left ( x \right ) = Cx+ D x \left (1-x \right )
\end{equation}
with $D=2Nd$.  Since for $D > 0$ the extra term $D x \left (1-x \right )$ favors coexistence
we expect that the coexistence phase, which for  $D=0$ is restricted to the surface $S=C/M$,  expands to occupy  a finite volume in the
space of parameters of the model. This is the reason in the following analysis we  will invest heavily on the analysis of the
regular steady-state solution of  eq.\ (\ref{eqeq}). Unless stated otherwise (see subsection \ref{negD}) we assume $D >0$.

%----------------------------------------------------------------------------
\subsection{Numerical analysis}
%----------------------------------------------------------------------------

Since $\hat{\phi_r} \left ( x \right )$ must be positive we write the regular solution of  eq.\ (\ref{eqeq}) in the form 
\begin{equation}\label{phir}
 \hat{\phi_r} \left ( x \right )=  \hat{\phi}_k \left ( x \right )  e^{y \left ( x \right )} 
\end{equation}
where $\hat{\phi}_k$ is  given by (\ref{sol}) and corresponds to the solution for the case
$R  = 0$ and $D=0$. In addition, in contrast to $ \hat{\phi_r}$, the function  $y$ is
always finite at the extremes $x=0$ and $x=1$. In terms of the auxiliary function 
$ z  = dy/dx $  we have
\begin{equation}\label{z}
x \left  (1-x \right ) \left ( z' + z^2 + S z   + D \right ) +  \left ( x-\bar{x} \right ) \left ( R - M z \right ) = \bar{D} 
\end{equation}
where
\begin{equation}\label{xbar}
\bar{x} = \int_0^1 dx \, x \hat{\phi}_k \left ( x \right ) e^{y \left ( x \right )}, 
\end{equation}
\begin{equation}\label{Dbar}
\bar{D} = D \int_0^1 dx \,x \left (1-x \right ) \hat{\phi}_k \left ( x \right ) e^{y \left ( x \right )},
\end{equation}
and
\begin{equation}\label{y}
y \left ( x \right )= y \left ( 0 \right )+ \int_0^x d\xi z \left ( \xi \right ) .
\end{equation}
Here the initial value $y \left ( 0 \right )$ is chosen 
in order to ensure the normalization, i.e. $ \int_0^1 dx \, \hat{\phi}_k \left ( x \right )  e^{y\left ( x \right )} =1$. 
We note that the values of $z \left ( x  \right )$ at the 
two extremes $x=0$ and $x=1$ are completely specified by eq.\ (\ref{z}),
\begin{equation}\label{z0}
z (0) = \frac{R}{M} + \frac{\bar{D}}{M\bar{x}} 
\end{equation}
and
\begin{equation}\label{z1}
z (1) = \frac{R}{M} - \frac{\bar{D}}{M(1-\bar{x})}  .
\end{equation}

At this stage the problem is ready for a numerical approach. For fixed $\bar{x}$ and $\bar{D}$ we solve 
eq.\ (\ref{z}) by propagating the Runge-Kutta algorithm from $x=0$ to $x=1$ using the initial condition 
(\ref{z0}). Of course, the choice of an arbitrary value of $\bar{D}$ will not satisfy the boundary condition (\ref{z1}) so we 
adjust $\bar{D}$ in order that condition is satisfied. This is essentially an application of the well-known shooting method to solve boundary values
problems \cite{Press_92}. Note that this
procedure actually accounts for replacing eq.\  (\ref{Dbar}) by the boundary condition (\ref{z1}).
Once this is achieved, we have solved the problem  for a fixed $\bar{x}$. We then calculate $\bar{x}$ using
(\ref{xbar}) and return to eq. (\ref{z}) repeating the process until  we reach the convergence for $\bar{x}$. 

%%%%%%%%%%%%%%%%%%%%%%%%%%
\begin{figure}[!t]
\begin{center}
\subfigure{\includegraphics[scale=0.75]{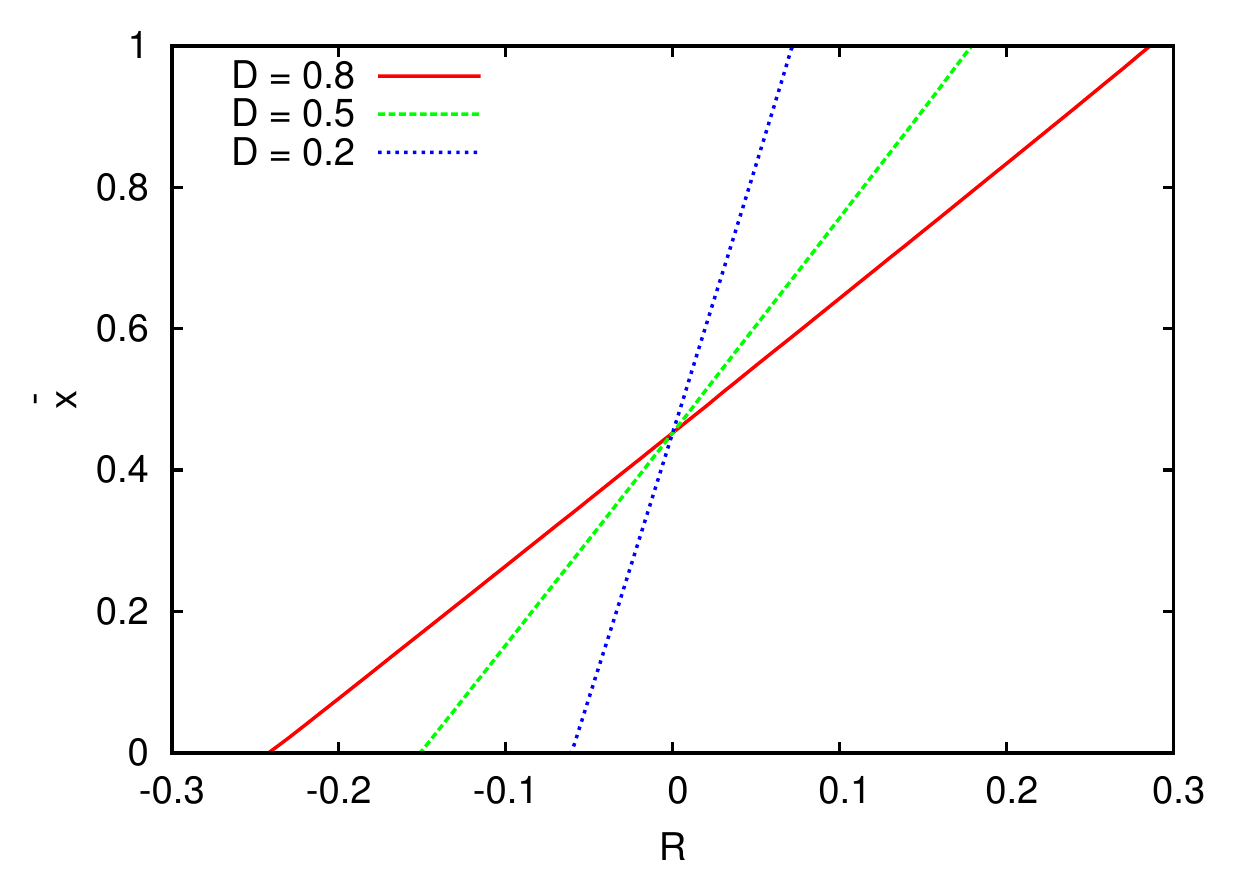}}
\subfigure{\includegraphics[scale=0.75]{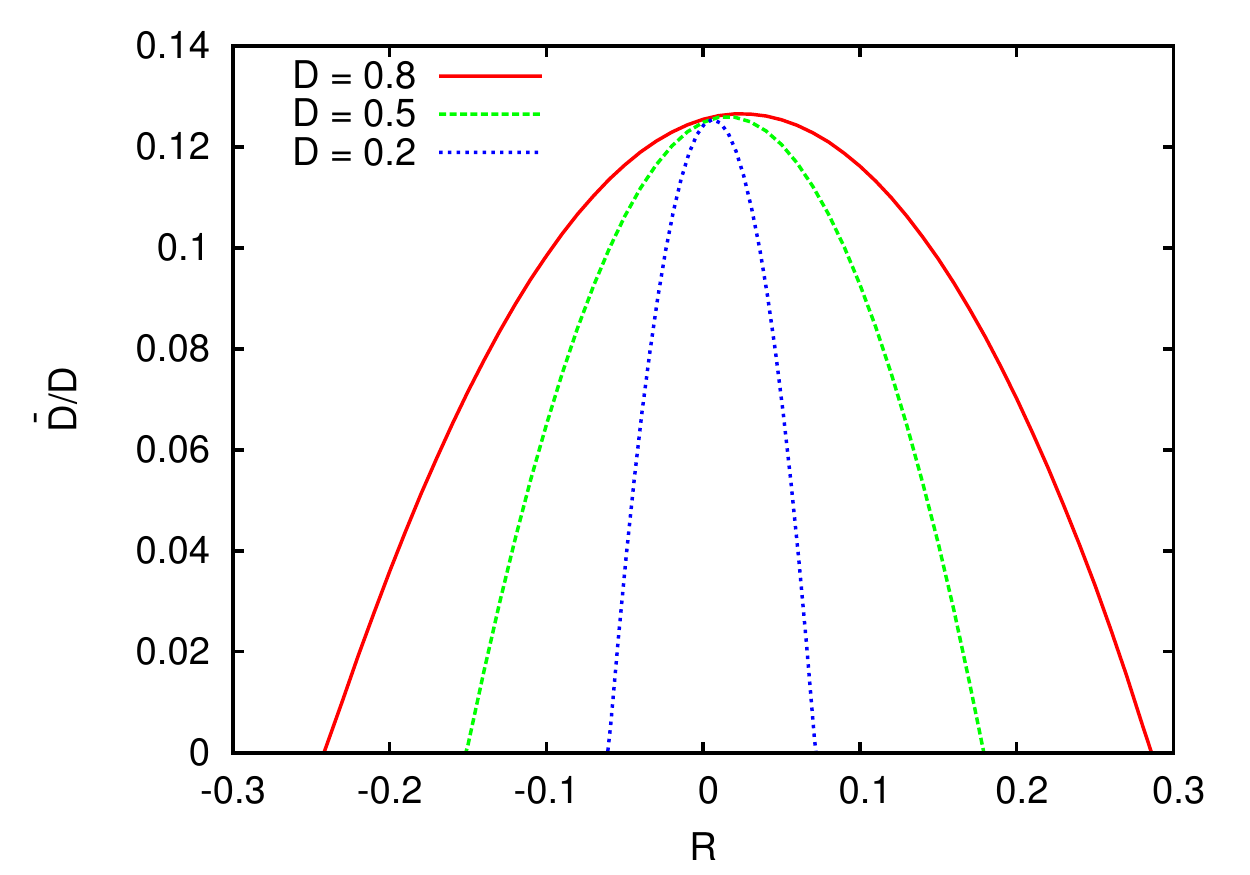}} 
\end{center}
\caption{Mean frequency of altruistic individuals $\bar{x}$ (upper panel) and coexistence index
$\bar{D}/D$ (lower panel) at the steady state as function of $R = C - MS$ for $C=1$,  $M=1$ and $D = 0.2, 0.5, 0.8$ as indicated in the
figure.  At $R=0$ we find $\bar{x} \approx 0.452$ and $\bar{D}/D \approx  0.124$. In the linear group survival case ($D=0$),
$\bar{x}$ exhibits a sharp transition from $0$ to $1$ at $R=0$,  at which its  value depends on the initial condition.} 
\label{fig:1}
\end{figure}
%%%%%%%%%%%%%%%%%%%%%%%%%%

Figure \ref{fig:1} summarizes the main results obtained using the above numerical scheme. In the coexistence phase, the  mean
frequency of altruists is well described by a straight line and  to a good approximation it seems to be independent  of
 $D$ for  $R=0$. The coexistence index $\bar{D}/D$ 
provides information on the mean balance of the coexistence within groups: it reaches the maximum value $1/4$   
for well-balanced groups, i.e., $\hat{\phi} = \delta \left ( x - 1/2 \right )$ and it vanishes outside the coexistence phase.

%----------------------------------------------------------------------------
\subsection{Transition lines}
%----------------------------------------------------------------------------

According to Fig.\ \ref{fig:1} we identify three phases in the steady-state regime: 
 the non-altruistic phase (NA) for which $\bar{x} = 0$, the coexistence  phase (C) for which
$0 < \bar{x} < 1$, and the altruistic phase (A) characterized by $\bar{x} = 1$.

The  transition line that separates phases NA and  C can be obtained by considering the limits
$\bar{x} \to 0$ and $\bar{D} \to 0$ of the regular solution $\hat{\phi}_r$. In this case eq.\  (\ref{z}) reduces to 
\begin{equation}\label{zxe}
 \left  (1-x \right ) \left ( z' + z^2 + S z   + D \right ) +  R - M z  = 0 
\end{equation}
that must be  solved using the boundary condition at $x=1$, eq.\ (\ref{z1}), which rewrites
\begin{equation}\label{ze}
z \left  (1 \right ) = \frac{R}{M}  .
\end{equation}
This boundary value problem  yields easily to a numerical approach (e.g., the shooting method \cite{Press_92}) 
which then allows us to obtain the function $z \left ( x \right)$ and, in particular, its value at the left boundary, $z \left ( 0 \right)$,
for arbitrary values of the parameters $S$, $M$, $C$ and $D$.
However,  since eq.\ (\ref{zxe}) is valid  at the transition line  only we need another condition
to  constraint the values of these parameters. This supplementary condition is  provided by  eq. (\ref{z0}) which,
after insertion of eqs.\ (\ref{xbar}) and (\ref{Dbar}),  reads
\begin{eqnarray}\label{criticale}
z(0)  & = & \frac{R}{M} +\frac{D}{M} \frac{\int_0^1 dx  
e^{y\left(x\right )} \left ( 1-x \right )^M }{\int_0^1 dx  
e^{y\left(x\right )} \left (1-x \right )^{M-1}} \nonumber \\
&  = & \frac{R}{M} +\frac{D}{M+1} \, \frac{\int_0^1 dx  
e^{y\left(x\right )} \rho_{M+1} (x)}{\int_0^1 dx  
e^{y\left(x\right )} \rho_M \left (x\right )}
\end{eqnarray}
where we have introduced the probability density
\begin{equation}\label{rho}
 \rho_M \left ( x \right ) = M \left (1-x \right )^{M-1} .
\end{equation}
Notice that the above expressions do not depend on the normalization factor $y(0)$, which actually diverges
in the limits $\bar{x} \to 0$ and $\bar{x} \to 1$.
The transition line is obtained by fixing $S$, $M$  and $C$ and adjusting  $D$ such that the value of  $z$ 
at the $x=0$ boundary of eq.\ (\ref{zxe}) coincides with the value obtained using expression (\ref{criticale}). 

Now we turn to to the  transition line that separates phases C and  A which  is obtained by considering the limits
$\bar{x} \to 1$ and $\bar{D} \to 0$ of the regular solution $\hat{\phi}_r$. In this case eq.\  (\ref{z}) reduces to 
\begin{equation}\label{zca}
x  \left ( z' + z^2 + S z   + D \right ) - R + Mz    = 0 
\end{equation}
which must be  solved  using the  boundary condition (\ref{z0}), i.e.,  
\begin{equation}\label{za}
z \left  (0 \right ) = \frac{R}{M} . 
\end{equation}
The procedure is identical to the sketched above for the transition line between the NA and C phases and so
the transition line is obtained by equating the value of $ z \left (1 \right ) $ that results from the solution of
the boundary value problem with the value  given by eq.\  (\ref{z1}),
\begin{eqnarray}\label{criticala}
z \left (1 \right ) & = & \frac{R}{M} - \frac{D}{M} \frac{\int_0^1 dx e^{y \left (x \right)}  
x^M}{\int_0^1 dx  e^{y \left (x \right)} 
x^{M-1}} \nonumber \\
&  = & \frac{R}{M} -\frac{D}{M+1} \, \frac{\int_0^1 dx  
e^{y \left( x \right)} \rho_{M+1}\left  (1-x \right )}{\int_0^1 dx  
e^{y \left( x \right)} \rho_M \left(1-x \right)}.
\end{eqnarray}
As before, the above expressions do not depend on the (divergent) normalization factor $y \left ( 0 \right )$.

\begin{figure}[!t]
\centering\includegraphics{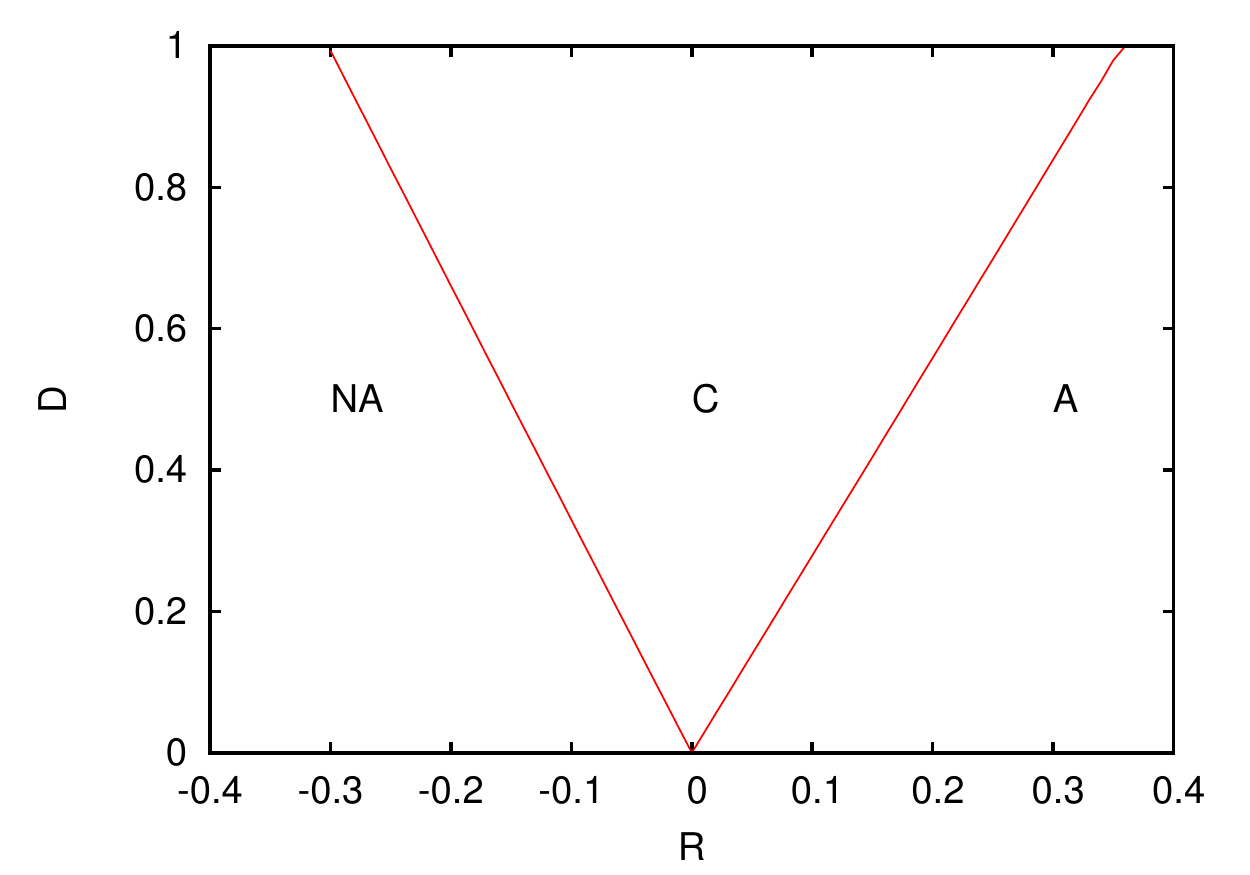}
\caption{Phase diagram of the model with the quadratic group survival function $C \left (x \right ) = C x + D x \left ( 1 - x \right) $
for $M=1$ and $C=1$ showing the 
 non-altruistic (NA),   ergodic coexistence (C) and altruistic (A) phases  in the space of parameters $D \leq C$ and $R = C - MS$.
In the linear group survival case (the  $D=0$ axis), the coexistence phase is limited to the point $R=0$
and it is non-ergodic.
}
\label{fig:2}
\end{figure}

Figure \ref{fig:2} exhibits the phase diagram of the model  for $C=1$ and $M=1$. The  transition lines are well 
fitted by straight lines (see Sect. \ref{sub:TA}) because of the constraint that $D < C$. In fact, allowing arbitrarily large values of
$D$  yields  significant deviation from those straight lines (data not shown).

%--------------------------------------------------------------------------------------
%--------------------------------------------------------------------------------------

%
%-----------------------------------------------------------
\subsection{Theoretical  analysis}\label{sub:TA}
%-----------------------------------------------------------
%

A remarkable feature of the phase diagram exhibited in Fig.\ \ref{fig:2} is that the transition lines are well-fitted
by straight lines within the region of interest, namely, $D \leq C$. This finding motivates the search for an analytical
solution of eq.\ (\ref{eqeq}) in the regime where the parameters $R =   C - SM $ and $D$ are small, i.e.,
close to the transition line of the linear problem (see Section \ref{sec:linear}).  The other parameter $S$, $M$  and $C$, however,
are not necessarily small.

The starting point of our approximation scheme is the identity 
\begin{equation}\label{crit}
R \int_0^1 e^{Sx}\hat{\phi} \left  ( x \right ) \left (x -\bar{x} \right ) \,dx 
+ \int_0^1 e^{Sx} \hat{\phi} \left (x \right ) \left [ D x \left (1-x \right ) -\bar{D} \right ] \,dx =0
\end{equation}
which is derived by multiplying  both sides of eq.\ (\ref{eqeq})
by $e^{S x}$ and then  integrating over  the interval $ \left [ 0,1 \right ]$.  Here 
 $\bar{x}$ and $\bar{D}$ are given by eqs.\  (\ref{barxc}) with
 $C\left ( x \right ) = C x + D x \left (1 - x \right ) $.
We note that eq.\  (\ref{crit}) is satisfied both by $\hat{\phi} \left ( x \right )  =\delta \left ( x - 1\right ) $ 
(phase A) and $\hat{\phi} \left ( x \right )  =\delta \left ( x \right )$ (phase NA). In addition, 
if there is coexistence, it must also be satisfied
by a regular function $\hat{\phi} \left ( x \right ) = \hat{\phi}_r \left ( x \right )$ (phase C).

Close to the transition point  $R=D=0$ in the coexistence phase (see phase diagram of Fig.\ \ref{fig:2}) 
we can  replace  $\hat{\phi}_r \left ( x \right )$ in eq.\ (\ref{crit}) with  $\hat{\phi}_k \left ( x \right )$, which is given by eq.\ (\ref{sol}).
In doing so we neglect  terms of second order on $D$ and  $R$.
The solution (\ref{sol}) has $\bar{x}$ as free parameter but close to  the transition lines
one must have $\bar{x} \to 0$ (transition from phase C to NA) and 
$\bar{x} \to 1$ (transition from phase C to A).

Let us consider  first the  case $\bar{x} \to 0$.  It can be easily verified that for any
arbitrary regular function $f \left ( x \right )$ we can write
\begin{equation}\label{f}
\int_0^1 f \left ( x \right ) \hat{\phi}_k \left  (x \right )\,dx \simeq
f \left  (0 \right ) + \bar{x} \int_0^1 \frac{f \left (x \right )-f \left  (0 \right )}{x} \rho_M  \left (x \right)\,dx 
\end{equation}
where $\rho_M$ is given by eq.\ (\ref{rho}) and we have neglected
 terms of higher order in $\bar{x}$. Note that the normalization condition ($f \left ( x \right ) = 1$)
and the mean ($f \left ( x \right ) =  x$) are preserved in this approximation scheme. The other moments
are correct to first order in $\bar{x}$. Hence, we can rewrite eq.\  (\ref{crit})  as
\begin{equation}\label{crits}
R \int_0^1 \left ( e^{Sx}-1 \right ) \rho_M \left ( x \right )\,dx 
+ \frac{DM}{M+1} \int_0^1 \left  (e^{Sx}-1 \right )\rho_{M+1} \left ( x \right ) \,dx =0
\end{equation}
which immediately gives the (positive) critical value $D_N$
at the transition line separating the coexistence and the non-altruistic phases
in terms of the (negative) parameter R,
\begin{equation}\label{coxE}
D_N= - \alpha_{N} R  
\end{equation}
where  $\alpha_{N}  = \alpha_{N} \left (C,M \right ) $ is given by
\begin{eqnarray}\label{alphaE}
\alpha_N & = & \frac{M+1}{M}
\frac{\int_0^1 (e^{Sx}-1)\rho_{M} \left ( x \right )\,dx }
{\int_0^1 (e^{Sx}-1)\rho_{M+1} \left ( x \right ) \,dx }  \nonumber \\
& = & \frac{1}{M} \frac{\sum_{n=0}^\infty  \zeta_{n+1} \left ( M \right ) \left ( C/M \right)^n}
{\sum_{n=0}^\infty  \zeta_{n+2} \left ( M \right ) \left ( C/M \right)^n}
\end{eqnarray}
where
\begin{equation}
\zeta_n \left ( M \right ) = \prod_{i=1}^n \frac{1}{M+i} ,
\end{equation}
and we have replaced $S$ by $C/M$ which is inconsequential to first order in $R$.
These calculations can be repeated in a completely analogous way to derive
the critical value $D_A$
at the transition line separating the coexistence and the altruistic phases. Recalling
that at this line we have $\bar{x} \to 1$, we find
\begin{equation}\label{coxA}
D_A=  \alpha_A  R 
\end{equation}
where $\alpha_{A}  = \alpha_{A} \left (C,M \right ) =  \alpha_{N} \left (-C,M \right )$.
For $C=M=1$,   eq.\ (\ref{alphaE}) yields  $\alpha_N = 3.29$  and  $\alpha_A= 2.78$, which match
perfectly  the slopes of the straight lines shown in the  phase diagram of Fig.\ \ref{fig:2}.
 Note that only
for $C=0$ (and hence $S=0$)  we  have  symmetry around the $R=0$ axis, i.e., $\alpha_{A}  = \alpha_N = \left ( M + 2 \right )/M$, and
in this case the coexistence phase is confined to the region
\begin{equation}\label{RegCoex}
D > \frac{M+2}{M} \mid R \mid .
\end{equation}

Now we set out to establish analytical approximations for the  mean coexistence group pressure   $\bar{D}$ 
and for the mean frequency of altruists $\bar{x}$ away from the transition lines. The  expression  for $\bar{D}$ 
 to first order in  $D$  follows immediately  from eq.\ (\ref{Dbar}),
\begin{equation}\label{d}
\bar{D}= D\int_0^1 \hat{\phi}_k \left (x \right ) \, x\, \left (1 -x \right ) \,dx = 
D \frac{M}{M+1}\bar{x}(1-\bar{x}) .
\end{equation}
However, the calculation of $\bar{x}$ to the leading  order in $R$ and $D$  is somewhat more involved.
We begin by noting that, according to Fig.\ \ref{fig:1}, the value of  $\bar{x}$ at $R=0$ appears
to be independent of  the parameter $D$. 
Alas,  by setting $R=0$ in eq.\ (\ref{crit}) we can see that this conclusion 
is not correct  since
the regular steady-state solution $\hat{\phi} = \hat{\phi}_r$ does depend on $D$. Next,
assuming  that $D$ is small we can use  eq.\ (\ref{d}) to eliminate $\bar{D}$  in eq.\ (\ref{crit}), which to
the lowest order in $R$ and $D$ is rewritten as 
\begin{equation}\label{xbRi}
\int_0^1 e^{Cx/M} \hat{\phi}_k \left (x \right ) \left [ \frac{R}{D} \left ( x - \bar{x} \right ) + 
 x \left ( 1 - x \right ) - \frac{M}{M+1} \bar{x} \left (1 - \bar{x} \right ) \right ] dx = 0 
\end{equation}
where we have replaced $S$ with $C/M$ and $\hat{\phi}$ with  the Beta distribution $\hat{\phi}_k$ given in (\ref{sol}).  
For the purpose of numerical evaluation
eq.\ (\ref{xbRi}) is rewritten as
\begin{equation}\label{xbR0}
\sum_{i=1}^\infty \frac{\left ( C/M \right )^i}{\left ( i-1 \right )!} \, \frac{ B \left ( \alpha  + i,  \beta \right )}{ M+i }
\left [ \frac{R}{D}  +   \frac{ \left ( M + 1 \right ) \beta  - \left ( M + i \right ) \alpha }{\left ( M +1 + i \right ) \left ( M + 1\right )}  \right ]
= 0
\end{equation}
with $\alpha = M \bar{x}$, $\beta = M \left ( 1 - \bar{x} \right )$ and $B \left (.,.\right )$ is the Beta function.
Solving  eq.\ (\ref{xbR0})  
yields $\bar{x}$ in terms of the parameter   $M$ and of the ratios $C/M$ and  $R/D$. In fact, this equation 
explains the observed, though not strict, lack of dependence  of $\bar{x} $ and $\bar{D}/D$ on the
parameter $D$ for $R=0$ (see Fig.\ \ref{fig:1}).  In particular, for $C=M=1$  and $R=0$ we find $\bar{x} = 0.453$ and,  inserting
this value in  eq.\ (\ref{d}),  $ \bar{D}/D = 0.124$,  which is
in good agreement with the results of Fig.\ \ref{fig:1}. In addition,
varying $R/D$  and solving eq.\ (\ref{xbR0}) for $\bar{x}$  yields results that are indistinguishable from  those  exhibited in the 
upper panel of Fig.\ \ref{fig:1}.  The coexistence indexes exhibited  in the  lower panel of Fig.\ \ref{fig:1} are equally very well described by inserting the
values of $\bar{x}$ into eq.\ (\ref{d}).  For $C \to 0$ (and hence $S \to 0$) eqs.\ (\ref{xbR0}) and (\ref{d}) yield
\begin{equation}\label{bxsmall}
\bar{x} = \frac{1}{2} +  \frac{\left ( M+2 \right ) R }{2MD} 
\end{equation}
and
\begin{equation}\label{bDsmall}
\bar{D}/D = \frac{M}{4 \left ( M + 1 \right )} \left [ 1 - \left ( 1 + \frac{2}{M} \right )^2   \frac{R^2 }{D^2} \right ] .
\end{equation}

We recall that the transition lines $ D_N = - \alpha_N \left ( C,M \right ) R$ and $ D_A = \alpha_A \left ( C,M \right ) R $
were derived by taking the limits $\bar{x} \to  0 $ and $\bar{x} \to  1 $, respectively, in eq.\ (\ref{xbRi}). Hence the values
of $\bar{x}$ obtained  by solving the clumsy eq.\ (\ref{xbR0}) or, equivalently,  eq.\ (\ref{xbRi}), tend to the correct limits at those transition lines.

We can derive a handy  approximation
for $\bar{x}$ with the aid of Fig.\ \ref{fig:1} by considering the equation of the 
straight line that joins the points $\left ( -D/\alpha_N, 0 \right )$ and $\left (D/\alpha_A , 1 \right )$, i.e.,
\begin{equation}\label{xbs}
\bar{x} = \frac{\alpha_A }{\alpha_A + \alpha_N} +  \frac{\alpha_A \alpha_N}{\alpha_A + \alpha_N}\frac{R}{D} 
\end{equation}
with $\alpha_N \left (C,M \right ) = \alpha_A \left ( -C,M \right) $ given by eq.\ (\ref{alphaE}). For $C=M=1$ this 
approximation scheme yields  $\bar{x}= \alpha_A/\left ( \alpha_A + \alpha_N \right) \approx 0.458$  at $R=0$, which is
very close to the result obtained by solving eq.\  (\ref{xbR0}) with $R=0$. 
 Note that the slopes of the approximate straight lines illustrated in the upper panel
of Fig.\ \ref{fig:1} are proportional to $1/D$ and therefore diverge at $D=0$.

%%%%%%%%%%%%%%%%%%%%%%%%%%%%%%%%%%%%%%%%%%%%%%%%%%%%

%----------------------------------------------------------------------------
\subsection{The case $D < 0$}\label{negD}
%----------------------------------------------------------------------------

Up to now we have considered the case where the nonlinear contribution to the
group survival rate explicitly favors coexistence, i.e, $ 0 < d < c$ (or, in terms of the rescaled parameters,
$0 < D < C$) in eq.\  (\ref{cx}). This choice amounts to saying that  the group survival $C \left ( x \right )$ is a non-decreasing
concave function of the frequency $x$ of altruists  in the group. Now we address briefly what happens when $C \left ( x \right )$  is
a non-decreasing convex function of $x$, which corresponds to the choice $0 < -D < C$.

Since a negative value of the parameter $D$  hinders coexistence by construction (see eq.\ (\ref{cx})) and since for
$D=0$ we find coexistence only at the transition line $R= C-MS=0$  (see Section \ref{sec:linear}) we expect the
coexistence phase to be  obliterated  for $D < 0$. In the Appendix we offer an analytical argument to support 
 this prospect. However, the effect of $D < 0$ goes beyond destroying the coexistence phase at $R=0$: it introduces
a new (non-ergodic) bistable regime which allows the reaching of the altruistic phase for $R < 0$ and the non-altruistic phase
for $R > 0$ provided $\mid D \mid$ is sufficiently large compared 
to $\mid R \mid $. 
The Appendix presents a proof of this result in the limit that the
parameters $C$, $S$ and $\mid D \mid $ are small. As the two steady-state phases
that exist for $D < 0$ are not described by a regular steady-state solution we lack the tools to
determine the boundaries of the  region of bistability in the parameter space. A promising approach is 
to introduce a small symmetric mutation rate $\nu$, as done by Kimura to study the linear case \cite{Kimura_83}, and
then extrapolate the results for $\nu \to 0$. 
However, because the transition lines obtained in the Appendix for $D < 0$
and small values of the model parameters  are identical to the analytical continuation to the region $D < 0$ of the 
transition lines $D_N$ and $D_A$ derived in subsection \ref{sub:TA},
we conjecture here that the region occupied by the bistable phase in the  half-plane $D< 0$ is  
the mirror  of the region occupied by  the ergodic coexistence phase in the  half-plane $D > 0$.

%----------------------------------------------------------------------------
\section{Heaviside group survival rate}\label{sec:Heav}
%----------------------------------------------------------------------------

%%%%%%%%%%%%%%%%%%%%%%%%%%
\begin{figure}[!t]
\centering\includegraphics{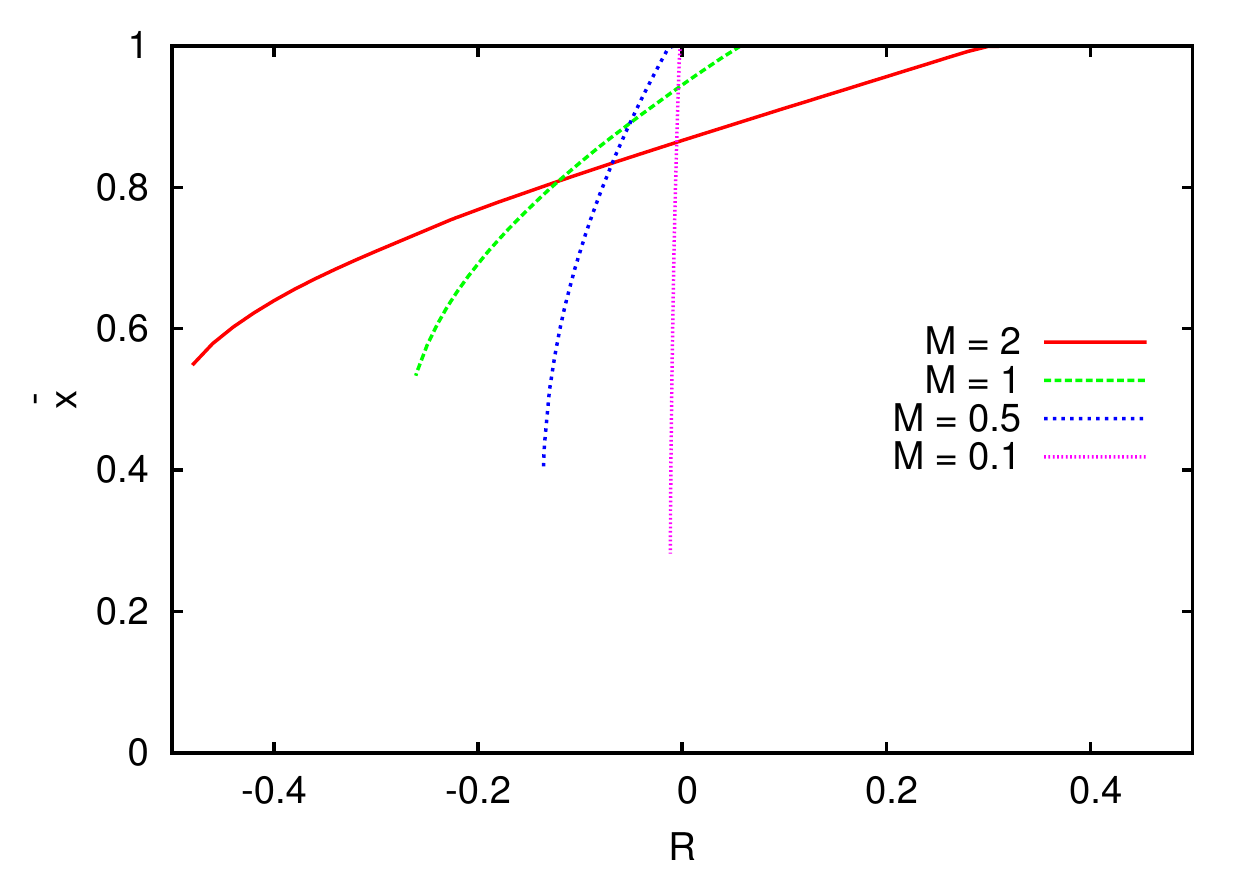}
\caption{Mean frequency of altruistic individuals $\bar{x}$ 
 as function of $R = C - MS$ for the Heaviside group survival rate. The parameters are $C=1$ and  $M$ as indicated in the
figure. }
\label{fig:3}
\end{figure}
%%%%%%%%%%%%%%%%%%%%%%%%%%

Here we consider a somewhat extreme  group selection pressure that sets off only in groups in which the
altruists are the majority of the group components, i.e.,
\begin{equation}\label{c_Heav}
  C \left ( x \right ) =  \left \{ \begin{array}{ll}
  							0 & \mbox{if \,\, $ 0 \leq x < 1/2$} \\
  							C & \mbox{if \,\,  $ 1/2 \leq x  \leq 1$. }
  						   \end{array}
  					\right.	   	
\end{equation}
This prescription models the division of labor between the altruists, termed synergism, and it
is useful to study the appearance of complex
structures that are of value to the organism only when fully formed \cite{Maynard_87,Alves_01}.
Using the  same transformations introduced in the previous section 
 we write the steady-state equation  (\ref{eqeq}) for the regular solution $\hat{\phi}_r = \hat{\phi}_k  \left ( x \right ) e^{y \left ( x \right )}$ as 
\begin{equation}\label{zHev1}
x \left  (1-x \right ) \left ( z' + z^2 + S z  \right )  - M  \left ( x-\bar{x} \right ) z  + C \left ( x \right ) = \bar{C} 
\end{equation}
where
\begin{equation}\label{CHbar}
\bar{C} = C \int_{1/2}^1 dx \;  \hat{\phi}_k \left ( x \right ) e^{y \left ( x \right )}.
\end{equation}
Here $\bar{x}$ and  $y \left ( x \right )$ are defined by eqs.\ (\ref{xbar}) and (\ref{y}) respectively. The boundary conditions
are $z \left ( 0 \right ) = \bar{C}/\left ( M \bar{x} \right )$ and $z \left ( 1 \right ) = \left ( C - \bar{C}\right )/\left [ M \left (1 - \bar{x} \right ) \right ]$.
As in the case of the quadratic group survival rate, this boundary value problem can be easily solved using the shooting method and Fig.\ \ref{fig:3} summarizes the main results. 
The noteworthy feature of this figure  that shows the dependence of $\bar{x}$ on $R=C - MS$  is the 
appearance of a discontinuous 
transition for negative $R$ that separates the non-altruistic and the coexistence phases. In particular, the jump in $\bar{x}$ at the transition line increases 
as the  migration rate $M$ increases. Moreover, for positive $R$  there is a continuous transition between the coexistence and the
altruistic phase. Whereas this continuous  transition can be located with good accuracy using the numerical approach of
the previous section as both $\bar{x}$ and $\bar{C}$ tend to $1$ at that transition, there is no shortcut to determine
the discontinuous transition as the values of those two variables are unknown  in this case. Since, in addition, we can offer no
analytical  support to the numerical results, we opt to restrict the study of the Heaviside group survival rate to  the
exhibition of Fig.\ \ref{fig:3} which proves our main point: a concave-like nonlinear   survival  rate,  such that
the  benefit for the  group increases slower and slower as the number of altruists increases,
favors the coexistence of the two types of individuals inside the group.

We note that provided   the  initial densities  are not   $\phi \left ( x,0 \right) = \delta \left ( x \right ) $ (i.e., only purely egoistic groups)
 or $\phi \left ( x,0 \right) = \delta \left ( x -1\right ) $ (i.e., only purely altruistic groups)
 the long term evolutionary dynamics will settle in the steady states described in Fig.\ \ref{fig:3} and, in particular, in  the coexistence regime for a 
proper choice of the model parameters $C$, $M$ and $S$. Even the linear combination 
$\phi \left ( x,0 \right) = a_0 \delta \left ( x \right ) + a_1 \delta \left ( x -1\right ) $ with $a_0 + a_1 = 1$
 can lead to coexistence in the long term. In this case, the migration process will play the key role by producing the mixed groups.  
In addition, if we start with an initial density such that  $x < 1/2$ for all groups so that the Heaviside  group selection (\ref{c_Heav}) is turned off, 
then we can invoke Haldane's argument to show that because  the groups have a finite size $N$ there is a non-vanishing probability of fixation of the altruists
in some groups  \cite{Haldane_32}, leading  back to the  abovementioned linear combination  of deltas in the worst case.

%-----------------------------------------------------------------------
\section{Conclusion}\label{sec:conc}
%-----------------------------------------------------------------------

Building on the diffusion model of group selection proposed
by Kimura \cite{Kimura_83},  in this paper we  offer an extensive study of the effects of the
convexity of the group survival function  $C \left ( x \right ) = C x + D x \left ( 1 - x \right )$ with $\mid D \mid \leq C$ 
on the steady-state properties of Kimura's model.  
As in the case that  the group survival rate increases linearly with the frequency $x$ of altruists
within  the group  \cite{Kimura_83},  we find that  a relevant independent variable in the resolution of the
phase diagram  of the model is the quantity  $R = C - M S$ where  
$M$ is the rescaled migration rate and $S$ is the rescaled selective advantage of the non-altruists. Typically, the
non-altruistic individuals dominate for $R$ large and negative, whereas the altruistic individuals dominate
for $R$ large and positive (see phase diagram of Fig.\ \ref{fig:2}). 
More pointedly, we find that the altruistic trait can be
maintained in the population provided that the condition 
\begin{equation} \label{conc}
R > - D/\alpha_N
\end{equation}
 is satisfied.
Here $D \geq 0$ and 
$\alpha_N > 0$ is given by eq.\ (\ref{alphaE}). Of course, condition (\ref{conc}) comprehends both the altruistic and
the coexistence phase.
 This condition is important  because it reveals that the exchange of individuals between
groups favors the non-altruistic trait, whereas in the case the groups are isolated (i.e., $M=0$) the altruistic trait 
prevails regardless of the selective advantage of the non-altruists, provided  either $C > 0$ or $D > 0$ \cite{Eshel_72,Aoki_82}.

It is interesting that in the case of the linear survival function the condition 
for the dominance of  the altruists $R > 0$ can be written as $ C \frac{1}{M} > S$, which is reminiscent of Hamilton's rule
\cite{Hamilton_64a,Hamilton_64b}  since $C$ can be interpreted as  the benefit accrued  to  all individuals in the group and $S$ as the cost (selective
disadvantage) paid by the altruists only. In Hamilton's rule the factor $\frac{1}{M}$ should be associated to the average relatedness of
the interacting individuals or, more generally, to some measure of the population structure \cite{Simon_12b}. This interpretation  holds true
in our case as well, since $M$ is proportional to the number of migrants and so the increase of $M$ results in the increase of 
interactions involving unrelated individuals, i.e., individuals coming from distinct groups.

We can get a  clue on the role of the group size $N$ by  reverting to
the original parameters $c = C/2N$, $d = D/2N$,  $m = M/2N$ and $s = S/2N$, so that the condition (\ref{conc}) for the 
sustenance of the altruistic trait becomes
 $c + d/\alpha_{N} > 2 N m s$.  This condition shows that the altruistic trait is favored if the groups are small enough
so that genetic  drift can fix the trait in a few groups, as pointed out  by Haldane long ago \cite{Haldane_32}. 
It is interesting to note that the  effective group size ranges from $N=10$ to $N=100$ for most 
vertebrate species \cite{Wilson_75}. However, that range increases vastly if one 
borrows the concepts of intergroup selection  to describe the evolution of parasite-host systems \cite{Levin_81}
and microbial populations \cite{Cremer_12}.
In that case, the hosts are associated with the groups and
the role of the altruistic individuals is played by the less virulent parasites which, by having a lower growth rate, 
increase the survival probability of the host.

In the case the survival rate is a concave function  of the frequency of altruists (i.e., $ 0 < D \leq C$)  we find an additional
phase -- an ergodic coexistence phase which monopolizes the region around $R = 0$, as illustrated in the phase diagram of
Fig.\ \ref{fig:2}. This finding contrasts with the linear case $D=0$ for which the non-ergodic coexistence phase occurs at $R=0$ only (see
Sect. \ref{sec:linear}).  The coexistence phase is separated from the altruistic and non-altruistic phases by two continuous transition lines,
which are very well-described by an approximation scheme based on the first order corrections to the solution of the  $D=R=0$ case.
In particular, we find that the average frequency of altruists in the meta-population $\bar{x}$, which can be viewed as the
order parameter of the model, vanishes or tends to unity linearly with the distance to the transition lines as those lines are approached from the
coexistence phase. Interestingly, these findings hold true for the case of a discontinuous  survival function (see Sect.\ \ref{sec:Heav}),
except that the transition between the non-altruistic and the coexistence phases becomes discontinuous.

In the case the survival rate is a convex function  of the frequency of altruists (i.e., $ 0 <- D \leq C$) the coexistence phase disappears
altogether, as expected. However, a new  non-ergodic phase  appears for values of $\mid D \mid$  large compared to $\mid R \mid $ 
in which either the altruistic or the non-altruistic phases can be  reached depending on the initial conditions. We conjecture that this 
bistable phase occupies a region in the half-plane $D < 0$ which is the reflection  over the $D=0$ axis of the region occupied by
the coexistence phase in the half-plane $D>0$. We stress that
coexistence is never allowed for $D$ negative.

The main result of this paper is that the region in the 
space of parameters where the altruist trait can be sustained in the population (see eq.\ (\ref{conc})) is enlarged significantly
if the group survival rate is a non-decreasing concave function of the frequency of altruists. In addition, we show the
utility of Kimura's formulation of intergroup selection based on the diffusion approximation of population genetics to produce
analytically treatable  two-level selection models.   Following the approach promoted by \cite{Simon_12a}, we gauge the
relevance  of group selection by the effects of  the group-level events (group extinction in our case) on the long term evolutionary dynamics.
In that sense, the existence of the coexistence and altruist regimes offers unequivocal evidence of the importance of group selection.

%-----------------------------------------------------------------------
\section*{Appendix}
%-----------------------------------------------------------------------

Here we present the calculations that unveil a nontrivial effect of negative
values of the parameter $D$, which amounts to a group pressure against
coexistence. 

In a similar manner  we derived  eq.\ (\ref{crit}), which is valid in the steady-state regime, we can
derive its dynamical counterpart by  multiplying both sides of equation (\ref{eqphi}) 
by $ e^{S x}$ and then integrating over $x$  over the  interval $ \left [0,1 \right ]$.
After rescaling  the time and  the model parameters we obtain
\begin{eqnarray}\label{critt}
\frac{d}{d \tau} \int_0^1 e^{Sx} \phi \left (x,\tau \right ) \,dx & = & 
R \int_0^1 e^{Sx}\phi \left (x,\tau \right )  \left  [ x -\bar{x}\left ( \tau \right ) \right ] \,dx  \nonumber \\
& & + \int_0^1 e^{Sx}\phi \left (x,\tau \right ) \left  [ D x \left (1-x \right ) -\bar{D}\left ( \tau \right ) \right ] \,dx .
\end{eqnarray}
We will consider the case that  $R = C - MS$,  $|D|$, and  $S$ (or $C$) are small 
so we can  keep only terms of the leading order on those parameters in eq.\ (\ref{critt}), 
yielding
\begin{equation}\label{critx}
\frac{d\bar{x} \left (\tau \right )}{d \tau} =  
 \int_0^1 x \left [  x -\bar{x}\left ( \tau \right ) \right ]   \left [ R + D \left (1-x \right ) \right ]
 \phi \left (x,\tau \right )  \,dx .
\end{equation}
 If there is a regular solution for the steady state, then
it must be, neglecting terms of higher order, the  Beta distribution (\ref{sol})
with a given $\bar{x}$.  Using
\begin{equation}
\int_0^1 x ^2 \hat{\phi}_k (x) \,dx = \bar{x} \left ( M\bar{x}+1 \right )/\left ( M+1 \right )
\end{equation}
and 
\begin{equation}
\int_0^1 x ^3 \hat{\phi}_k (x)\,dx = \bar{x}
\left ( M\bar{x}+2 \right ) \left ( M\bar{x}+1 \right )/\left [ \left (M+2 \right ) \left (M+1 \right ) \right ]
\end{equation}
we have
\begin{equation}\label{cxap}
\frac{d\bar{x} \left ( \tau \right )}{d \tau} =  \frac{ \bar{x} \left ( \tau \right ) \left [ 1-\bar{x} \left ( \tau \right ) \right ]}{M+1}
\left [ R + D \frac{M}{M+2} \left [1-2\bar{x}\left ( \tau \right ) \right ]
\right]
\end{equation}
whose stationary solution ($d\bar{x}/d\tau=0$) is given by 
\begin{equation}\label{again}
\bar{x} = \frac{1}{2} +  \frac{\left ( M+2 \right ) R }{2MD} ,
\end{equation}
which is identical to eq.\ (\ref{bxsmall}). Now, linearization  of eq.\ (\ref{cxap})
around $\bar{x}$ yields
\begin{equation}\label{deltat}
\frac{d\delta \left ( \tau \right )}{d \tau} = -D \frac{2M}{M+2}
\frac{\bar{x}(1-\bar{x})}{M+1} \delta \left ( \tau \right )
\end{equation}
where $\delta \left ( \tau \right ) = \bar{x}\left ( \tau \right ) -\bar{x} \ll 1$ as usual.
Therefore  solution (\ref{again}) is stable for positive $D$
and unstable for negative $D$.  

Let us focus on the case $D<0$  only. In this case the unstable fixed point
(\ref{again}) exists provided that the condition 
\begin{equation}\label{nonerg}
\mid D \mid \geq  \frac{M+2}{M} \mid R \mid
\end{equation}
is satisfied.
Furthermore, if this condition is satisfied  then it can be easily proved that 
the right hand side of eq.\  (\ref{cxap}) does not change sign
during evolution, so that the final value for $\bar{x}$ is either $0$ or $1$ 
depending on the initial condition.  Since this bistable phase
depends on the existence of the unstable fixed point (\ref{again}) and that 
this fixed point tends to $0$ or $1$ when the equality holds in condition (\ref{nonerg}) then
the transitions between this phase and the two other ergodic  phases are continuous.
Most importantly, the condition (\ref{nonerg})  for  the existence of
the bistable phase  in the case $D < 0$ is identical to the condition for the existence of
the ergodic coexistence phase in  the case $D > 0$, given by eq.\ (\ref{RegCoex}) for small $S$, $R$ and $D$.
Hence, we conjecture that the region occupied by the bistable phase is  exactly the reflection over the 
$D=0$ axis of the region occupied  by the ergodic coexistence phase.

For  $R=0$ we  can offer an alternative argument to show that the non-ergodic coexistence
regime discussed in Sect.\ \ref{sec:linear} is destabilized by the parameter $D < 0$. In this case
eq.\ (\ref{critt}) rewrites as
\begin{equation}\label{cov}
\frac{d}{d \tau} \int_0^1 e^{Sx} \phi \left (x,\tau \right ) \,dx = - \mid  D \mid \, {\rm Cov} \left [ e^{Sx}, x \left (1-x \right ) \right ]
\end{equation}
where ${\rm Cov} \left [ e^{Sx}, x \left (1-x \right ) \right ]$ is the covariance
between $e^{Sx}$ and $x \left (1-x \right )$.
Now, assume that $\bar{x} \left (0 \right )$ is close to 0, or, equivalently,
that $\phi \left (x,0 \right )$ is concentrated on $x= 0$.
In this case, the covariance is positive since  in the region of small 
$x$, where the density is more concentrated,  both
$e^{Sx}$ and $x \left (1-x \right )$ are increasing functions of $x$. Hence 
the integral $\int_0^1 e^{Sx} \phi  \left (x,\tau \right ) \,dx$
decreases with time which  implies that $\bar{x}  \left (\tau \right )$ decreases and $\phi (x,t)$
becomes more and more concentrated on $x=0$.
Therefore, the covariance remains positive and  for large $ \tau$ the integral
$\int_0^1 e^{Sx} \phi  \left (x,\tau \right )\,dx$ reaches its minimal value 1, which 
implies that $\bar{x} \left ( \tau \right )$ vanishes.
Parallel results are  derived when $\bar{x} \left ( 0 \right )$ is close to 1 and so the
covariance in eq.\ (\ref{cov}) is negative. In this case, $\bar{x} \left ( \tau \right )$ increases
until it reaches the value 1.

In summary, the  pressure against coexistence  associated to negative values of the parameter $D$
destroys the coexistence regime altogether, as expected. However, it introduces an unexpected non-ergodic
phase where the final outcome of the evolutionary dynamics  is the altruistic regime ($\bar{x}=1$) or
the non-altruistic one ($\bar{x}=0$) depending on the initial conditions. For small values of
the model parameters we find that this non-ergodic phase exists in the region (\ref{nonerg}). Outside
this region (i.e., for $\mid D \mid $ small compared to $\mid R \mid$) we have the ergodic altruistic phase for 
$R > 0$ and the ergodic non-altruistic phase for
$R<0$. Since none of these phases can be described by a regular steady-state solution, our analysis is
limited to the approximation scheme presented in this appendix.

\section*{Acknowledgments}
The work of J.F.F. was supported in part by Conselho Nacional de
Desenvolvimento Cient{\'\i}fico e Tecnol{\'o}gico (CNPq)  and  M.S. was 
partially supported by PRIN 2009 protocollo n. 2009TA2595.02.

\end{document}